# Simulations of Vortex Evolution and Phase Slip in Oscillatory Potential Flow of the Superfluid Component of $^4$He Through an Aperture

J. A. Flaten,[1] C. T. Borden,[2] C. A. Lindensmith,[3] and W. Zimmermann, Jr.[4]
[1]*Department of Physics, Luther College, Decorah, IA 52101, USA*
[2]*Michigan Technological University, Houghton, MI 49931, USA*
[3]*Jet Propulsion Laboratory, Pasadena, CA 91109, USA*
[4]*Tate Laboratory of Physics, University of Minnesota, Minneapolis, MN 55455, USA*

**Abstract.** The evolution of semicircular quantum vortex loops in oscillating potential flow emerging from an aperture is simulated in some highly symmetrical cases. As the frequency of potential flow oscillation increases, vortex loops that are evolving so as eventually to cross all of the streamlines of potential flow are drawn back toward the aperture when the flow reverses. As a result, the escape size of the vortex loops, and hence the net energy transferred from potential flow to vortex flow in such $2\pi$ phase-slip events, decreases as the oscillation frequency increases. Above some aperture-dependent and flow-dependent threshold frequency, vortex loops are drawn back into the aperture. Simulations are preformed using both radial potential flow and oblate-spheroidal potential flow.

## I. INTRODUCTION

When superfluid $^4$He potential flow through an aperture reaches the critical velocity associated with that aperture, energy begins to be dissipated from the flow. This energy dissipation is believed to be due to the motion of quantum vortices in the vicinity of the aperture. When a vortex core evolves in such a way as to cross streamlines of the potential flow through the aperture, energy is transferred from the potential flow field to the flow field of the vortex itself. The vortex, whose core may be in the shape of a closed ring or a loop with ends at the walls in or near the aperture, may then carry this energy away from the vicinity of the aperture by means of a combination of self-induced velocity and background potential flow.[1-4]

The amount of energy transferred to a quantum vortex is proportional to the mass current of the potential flow crossed by the vortex core as it evolves. Every vortex that evolves so as to cut all of the streamlines of a given constant potential flow through an aperture will transfer the same amount of energy $\kappa I_s$ from potential flow to vortex flow. Here $\kappa$ is the quantum of circulation, equal to $h/m_4$, where $h$ is Planck's constant and $m_4$ is the mass of the $^4$He atom, and $I_s$ is the superfluid mass current. Such vortex-crossing events are known as $2\pi$ phase slips because they result in a 2π change in the difference of the phase of the superfluid order parameter between the two reservoirs connected by the aperture.[4] Various sensitive experiments using fluid-dynamical resonators have shown series of abrupt, nearly identically-sized energy-loss events, consistent with individual, independent 2π phase slips involving vortices created by thermal or quantum nucleation.[5-14]

Several candidate geometries have been proposed in which quantum vortices will spontaneously cross all of the streamlines of potential flow through an aperture as the vortices evolve. These proposals involve the evolution of fairly simple vortex configurations in steady, unidirectional potential flow through symmetrical apertures. In particular, computer simulations have shown that vortex rings or loops of appropriate shape nucleated in the high-velocity flow region near a circular aperture can indeed evolve in such a way as to cross all of the lines of potential flow through the aperture.[2,3,15-19]

In experiments to study this phenomenon using fluid-filled resonators, the actual potential flow through the aperture has been oscillatory. (See, for example, Refs. 5-14, 20, and 21.) If the period of oscillation is long compared to the time that it takes for a vortex to cross all of the lines of potential flow, the flow may be approximated by steady potential flow. However, if this is not the case, the motion of a vortex and the energy that it carries away will be influenced by the time variation of the potential flow. During intervals of reversed flow, the flow will attempt to pull the vortex back toward the aperture, and at frequencies above some aperture-dependent and flow-dependent threshold value, a vortex may in fact be drawn all the way back into the aperture.

Three different experiments have been performed in Minneapolis in which the steady flow approximation may not have been valid, at least under some of the conditions studied. All of these experiments showed effects that may be related to the influence of oscillatory flow on the motion of vortex loops. In two experiments, certain changes in the dissipation involving large energy-loss events arose with increasing frequency of oscillation.[14,22-24] In the third experiment, energy losses less than the expected steady-flow values were recorded.[13]

In an attempt to begin to understand the vortex dynamics responsible for these effects, we have used computer simulations to explore vortex loop evolution in oscillatory potential flow through an aperture in some very simple situations. This article describes the results of these simulations. In this work, the initial formation of a vortex by thermal nucleation or quantum tunneling is not considered, nor is the early fluid-dynamical growth of the vortex. Rather, the simulations deal with the main period of vortex growth, during which almost all of the energy transfer from potential flow to vortex flow takes place and during which frequency-dependent effects occur.

## II. SIMULATIONS

Two computer simulation programs were used to study vortex loop evolution in steady and oscillatory potential flow through an aperture. Both simulations extend previous studies of vortex evolution in steady potential flow alone.[3,15-19] The first program, discussed in Sections II A and II B, simulates the motion of a semicircular vortex loop above a flat surface in radial potential flow emanating from a very small aperture in that surface. The second program, discussed in the Sections II C and II D, simulates the motion of a vortex loop in an oblate-spheroidal potential flow field, the necked boundary of which constitutes the aperture.

The following simplifications and idealizations are made in both simulations. A small semicircular vortex loop is started near or within the aperture and allowed to evolve according to the rules of classical fluid dynamics for inviscid, incompressible fluids. The vortex is released when the potential flow velocity through the aperture is at its maximum, the time at which vortex nucleation or depinning is most likely to occur. The potential flow velocity is assumed to vary sinusoidally in time, and vortex evolution is studied as a function of the frequency of oscillation. The amplitude of the potential flow is not adjusted to compensate for the energy transferred to the evolving vortex, as if the amount of energy dissipated in a single $2\pi$ phase slip is small compared to the maximum kinetic energy of the potential flow, a condition that was well satisfied in the Minneapolis experiments.

In each simulation run, only one vortex loop is present. The ends of the vortex loop are assumed to move freely over the surface of the aperture so that the vortex core meets the surface perpendicularly. No pinning is assumed to occur. Furthermore, no frictional effects are included. Thus the results correspond to vortex motion at temperatures below about 1 K, where there is essentially no normal fluid. In all cases studied, nonadjacent portions of the vortex core remain well separated, and except for the sections at the ends, the vortex core remains well away from the walls. Thus vortex reconnections need not be considered.

### A. Simulation with Radial Potential Flow

In the first simulation, oscillatory potential flow is assumed to radiate into the upper half-space from an origin in a horizontal flat plate, as if the flow were passing through a tiny aperture centered at the origin. Maximum upward flow is assumed to occur at time $t = 0$. Hence the potential flow velocity $v_p(\mathbf{r},t)$ at position $\mathbf{r}$ relative to the origin and time $t$ is given by

$$v_p(\mathbf{r},t) = \frac{Q_0}{2\pi} \frac{\mathbf{r}}{r^3} \cos(2\pi f t), \tag{1}$$

where $Q_0$ denotes the amplitude of the volume rate of flow away from the origin and $f$ is the frequency of the flow. At time $t = 0$ a semicircular vortex loop is assumed to be situated above the plate, with its axis lying in the plane of the plate and passing through the aperture, as shown in Fig. 1(a). The vortex loop is oriented so that its self-induced velocity will carry it toward the aperture.

When combined with its half-ring image in the plate, such a vortex loop has the same self-induced velocity as that of a complete circular vortex ring of the same radius in an unbounded fluid, which may be written as

$$v_v(R) = \frac{\kappa}{4\pi R} ln\left(\frac{8R}{e^{1/2} a_0}\right), \tag{2}$$

where $R$ is the radius of the vortex ring and $a_0$ is the vortex core radius parameter, equal to $1.3 \times 10^{-10}$ m.[25,26] The net motion of the vortex core is dictated by the vector sum $\mathbf{v}_p + \mathbf{v}_v$ at each point along its length.

Under these simple conditions, the vortex loop remains a perfect semicircle perpendicular to the plate for its entire evolution, and its center moves along a fixed axis, which we call the $x$-axis, that coincides with the axis of the initial vortex, as shown in Fig. 1. The symmetry of the situation greatly reduces the amount of calculation needed to determine the trajectory of any single point on the vortex loop. Such a trajectory may be determined by solving the following pair of first-order differential equations numerically,

$$\frac{dR}{dt} = \frac{Q_0 R \cos(2\pi ft)}{2\pi (R^2 + x^2)^{3/2}} \tag{3}$$

$$\frac{dx}{dt} = \frac{Q_0 x \cos(2\pi ft)}{2\pi (R^2 + x^2)^{3/2}} - \frac{\kappa \eta}{4\pi R}, \tag{4}$$

where $\eta = ln(8R / e^{1/2} a_0)$. Here $x$ is the position of the center of the loop along the $x$-axis relative to the origin at the aperture. The value of $x$ is taken to be positive at $t = 0$. The variable $R$ is the perpendicular distance of any point on the vortex core from the $x$-axis. For later use, we take the axis perpendicular to the plate passing through the aperture to be the $z$-axis, positive in the upward direction. These equations are integrated forward in time numerically using a simple Euler method.

For steady potential flow, if the initial self-induced velocity of the vortex loop is less than the x-component of the potential flow velocity at its core, the loop will initially be swept away from the aperture by the potential flow and expanded by it, as shown in Fig. 1(a). Eventually the self-induced velocity of the vortex, which decreases approximately as $1/R$ as the half-ring grows, will exceed the x-component of the potential flow field, which falls off as $1/r^2$ away from the origin. As a result, the vortex loop reverses its motion in the x-direction, still continuing to grow because of the radial potential flow. Eventually the vortex crosses over the origin, as shown in Fig. 1(b), and continues off to infinity. As it escapes, the vortex ultimately cuts all of the streamlines of potential flow emanating from the origin (if those initially passing through the loop are ignored). This process constitutes a $2\pi$ phase slip.

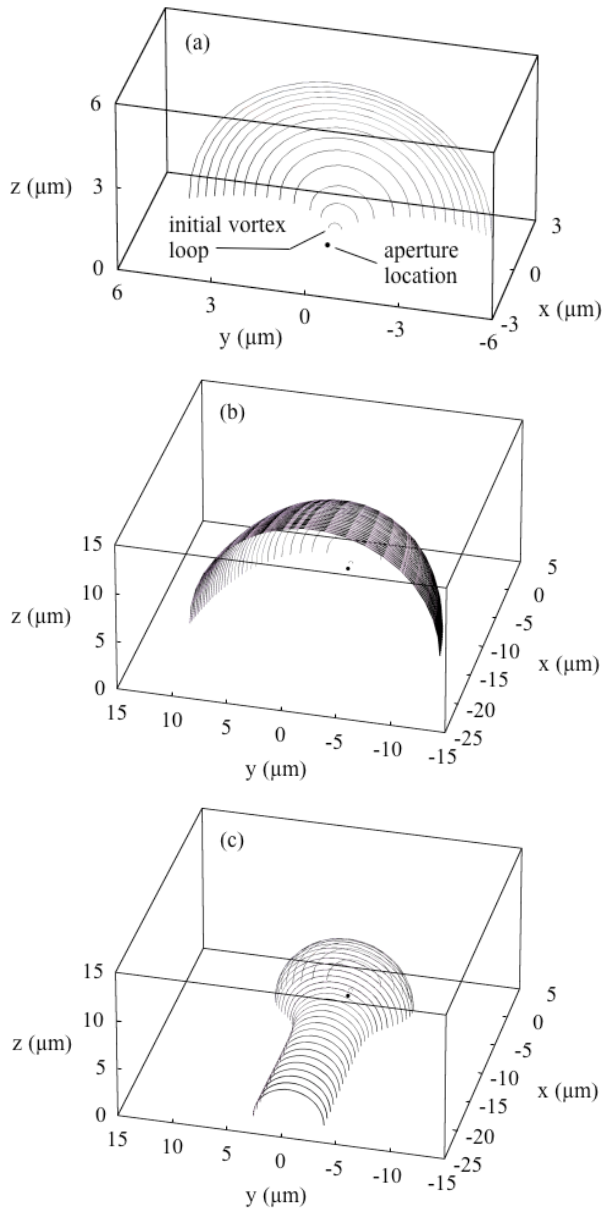

FIG. 1. Oblique views of a semicircular vortex loop evolving so as to cross the radial potential flow field emanating from a tiny aperture. The aperture, indicated by the dot at the origin, lies in the plane of the floor of the frame. (a) The early stage of vortex loop evolution in steady upward flow. The smallest loop shown is the initial vortex in the simulation. The vortex is first swept away from the aperture in the positive x-direction by the potential flow and expanded by it, and then begins to move in the negative x-direction. (b) Further vortex loop evolution in steady potential flow, showing how the vortex continues to grow and crosses over the aperture as it moves off toward $x = -\infty$. The first four loops in (b) correspond to the 1st, 6th, 10th, and last loops in (a). Throughout its motion, the vortex cuts lines of potential flow. (c) Vortex evolution in oscillatory potential flow at 800 Hz, in which the vortex loop shrinks in size when the flow reverses, ultimately escaping with a smaller radius than in case (b) and hence carrying away less energy. The total time durations of the plots in (b) and (c) are the same, equal to one full period of potential flow oscillation at 800 Hz.

A similar description of vortex evolution can be given in the case of oscillatory potential flow through the aperture. In this case the vortex is released when the potential flow is maximum, so the potential flow velocity first decreases as time elapses. This allows the evolving half-ring to reverse direction earlier than in the steady-flow case. Having a smaller radius, the vortex passes over the origin more quickly in the oscillatory case than in the steady case. After one quarter of a cycle, the potential flow velocity passes through zero and then increases in the opposite direction. When this happens, the flow attempts to draw the vortex back toward the origin, further decreasing its radius and increasing its self-induced velocity relative to the steady-flow case. Thus if the evolving vortex manages to cross the aperture and escape, it will do so with a smaller ultimate size than in the steady case, as shown in Fig. 1(c). This means that less energy is ultimately transferred from potential flow to vortex flow by $2\pi$ phase-slip events in oscillatory potential flow than in steady potential flow. The possibility also exists that the vortex will not escape before being drawn back into the aperture.

### B. Results from the Radial Potential Flow Simulation

Figure 2 illustrates the trajectories $z(x)$ of the peak of a semicircular vortex loop (actually, $R(x)$ for any point of the loop) evolving in steady potential flow, as well as in oscillatory potential flow at a series of frequencies $f$. The arrows and zeros on the traces indicate how far the vortex has evolved at successive upward maxima, zeros, and downward maxima (i.e. every quarter cycle) of the oscillatory flow. Eventually the vortex travels far enough away from the aperture so that the oscillation of the flow no longer causes the vortex radius to fluctuate significantly. After that, the vortex proceeds (to the left in Fig. 2) at a more-and-more nearly constant rate due to its self-induced velocity alone.

For Fig. 2 we have set $Q_0$ to be $4.0 \times 10^{-12}$ m$^3$/s and assumed initial values of $R = 0.2$ μm and $x = 1.0$ μm. With these choices, the initial self-induced velocity of the vortex is only 56% of the initial $x$-component of the potential flow velocity at the vortex core, so the vortex is initially swept away from the aperture. The motivation for these parameter choices is given in the conclusions section, where the relation to experiment is discussed.

As the frequency $f$ of the oscillatory potential flow increases, the vortex loop is closer and closer to the aperture when the flow reverses, and hence the trajectory of the loop is influenced more and more dramatically by the reversal. The ultimate size of a semicircular vortex loop that escapes decreases monotonically with increasing frequency. There exists a frequency threshold, near 1250 Hz for the parameters used here, above which half-rings no longer escape but are drawn back into the aperture. The trajectory of a vortex that is unable to escape is illustrated in Fig. 2 by the trace labeled 1400 Hz. As a general rule, the vortex loop must be somewhat more than half way across the aperture when the flow first reverses for it to escape. Because of the singularity in the potential flow-field at the origin, this simulation does not address the question of what happens to a vortex that is drawn back into the aperture after failing to escape.

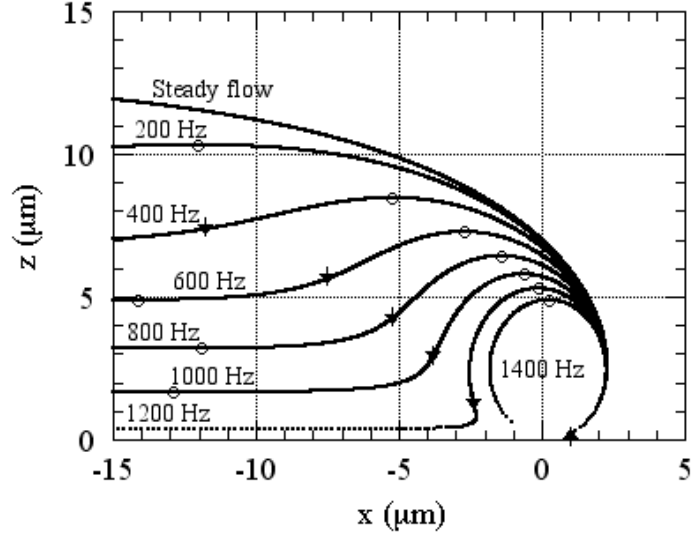

FIG. 2. Trajectories of the peak of a semicircular vortex loop evolving in steady and oscillatory radial potential flow. The potential flow initially emanates upward from a tiny aperture at the origin. In the case of oscillatory potential flow, up (and down) arrows indicate the location of the vortex peak at the upward (and downward) maxima of the flow. The zero-symbols indicate the location of the vortex peak when the flow goes through zero as it switches direction. Notice how at 1400 Hz the evolving vortex loop fails to escape but is instead drawn back to the aperture when the potential flow reverses.

Consistent with Eq. (2), the kinetic energy $E_v$ of a semicircular vortex loop at low temperatures is given by

$$E_v = \frac{\rho \kappa^2 R}{4} \ln\left(\frac{8R}{e^{3/2} a_0}\right), \quad (5)$$

where $\rho$ is the density of the fluid.[25,26] As the vortex grows, it extracts energy from the potential flow. However, when the flow reverses and the vortex shrinks again, energy is transferred from vortex flow back into potential flow. The net amount of energy ultimately transferred from potential flow to vortex flow during each $2\pi$ phase slip may be calculated by subtracting the initial vortex kinetic energy, which is nearly zero for the tiny initial half-ring used here, from the energy of the vortex when it reaches its asymptotic escape size.

Figure 3 shows a plot of the net amount of energy ultimately transferred from potential flow to vortex flow as a function of the frequency $f$ of potential-flow oscillation for the same parameters as used in Fig. 2. The energy transferred is normalized by the energy-transfer value for steady flow, the case in which the most energy is transferred. If the energy of the initial vortex can be neglected, as it can be here, the energy transferred for steady flow equals $\rho \kappa Q_0$.[1-3,5,20] As $f$ increases toward the threshold frequency for recapture, the escaping vortex carries away less and less energy. When the escaping vortex has negligible size but still escapes, just below the threshold frequency, a $2\pi$ phase slip occurs in which no net energy has ultimately been transferred from potential flow to vortex flow. Such an event might be described as a "dissipationless $2\pi$ phase slip".[27]

These results may be applied, at least approximately, to other values of $Q_0$ by recognizing that Eqs. (3) and (4) can be reduced to nearly $Q_0$-independent form by the introduction of dimensionless variables. One convenient procedure is to express all lengths except $a_0$ in units of $4\pi Q_0/\kappa$, to express time in units of $(4\pi)^3 Q_0^2/\kappa^3$, and to express frequency in units of $\kappa^3/\left((4\pi)^3 Q_0^2\right)$. The resulting equations then depend only weakly on $Q_0$, $\kappa$, and $a_0$ through a term $\ln(Q_0/\kappa a_0)$. To the degree that this dependence can be neglected, one can apply the results of Figs. 2 and 3 to a new value of $Q_0$ simply by multiplying all of the lengths except $a_0$ by $Q_{0\,new}/Q_{0\,old}$ and dividing all of the frequencies by $Q_{0\,new}^2/Q_{0\,old}^2$.

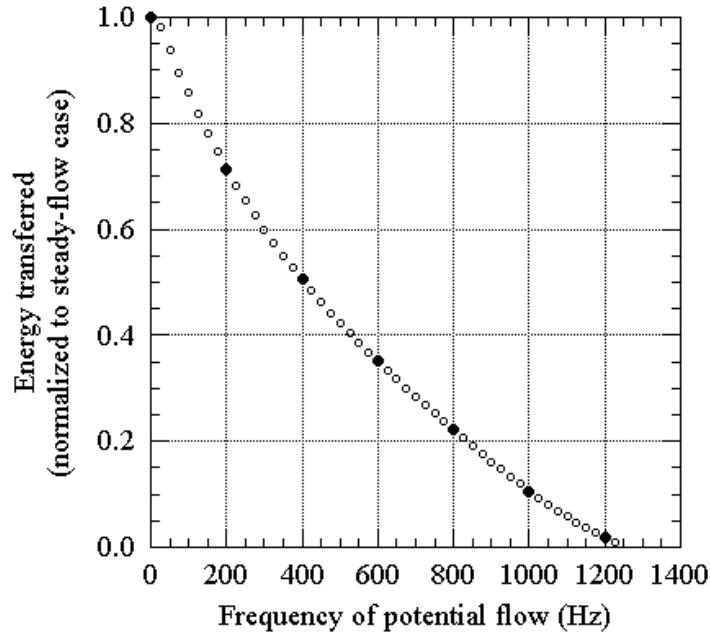

FIG. 3. The amount of energy ultimately transferred from potential flow to vortex flow by individual $2\pi$ phase-slip events such as those depicted in Fig. 2, as a function of the oscillation frequency of the potential flow. All energies are normalized to the value of the energy-transfer for steady flow. The initial energy of the vortex loop is negligible on this scale. Filled dots correspond to frequencies for which trajectories are shown explicitly in Fig. 2.

## C. Simulation with Oblate-Spheroidal Potential Flow

The second simulation program used was inspired by the work of Schwarz, who studied the evolution of a variety of vortex configurations in an aperture using an oblate-spheroidal steady potential flow field.[17-19,28] The relationship between Cartesian coordinates $x$, $y$, $z$ and oblate-spheroidal coordinates $\phi$, $\xi$, $\eta$ is

$$x = a\cos(\phi)\left[(\xi^2+1)(1-\eta^2)\right]^{1/2}$$
$$y = a\sin(\phi)\left[(\xi^2+1)(1-\eta^2)\right]^{1/2} \quad (6)$$
$$z = a\xi\eta,$$

where $a$ is a constant and $0 < \eta < 1$. Consider a region bounded by a surface of constant $\eta = \eta_0$, which defines a hyperboloid of revolution. If $\eta_0$ is not small compared to unity, the region $\eta_0 < \eta < 1$ forms a constricted channel whose axis coincides with the $z$ axis. (See Figs. 4, 5, and 6(a), where $\eta_0 = 0.50$.) If $\eta_0$ is small compared to unity, this region resembles a circular aperture in a wall. (See Fig. 6(b), where $\eta_0 = 0.05$.) The minimum radius $b$ of the channel or aperture varies with $\eta_0$ according to

$$b = a\left(1-\eta_0^2\right)^{1/2}. \quad (7)$$

When an ideal fluid flows through this channel or aperture, the magnitude of the potential flow velocity $v(\phi,\xi,\eta)$ is related to the maximum flow velocity $v_c$ at the center of the aperture by the expression

$$v(\phi,\xi,\eta) = v_c\left[(\xi^2+\eta^2)(\xi^2+1)\right]^{1/2}. \quad (8)$$

The total volume rate of flow through the aperture $Q$ is given in terms of $v_c$ by the expression

$$Q = 2\pi a^2(1-\eta_0)v_c. \quad (9)$$

Schwarz studied the evolution of a small vortex loop in steady potential flow through an oblate spheroidal channel. He showed that an appropriately-sized vortex loop in the form of a circular arc, started in the constricted region of the channel with its self-induced velocity pointed upstream, could be swept downstream out of the channel by the potential flow and evolve so as ultimately to cross all of the streamlines of the potential flow. This process constitutes a $2\pi$ phase slip in a way that is qualitatively similar to the process considered in Sections II A and II B.

In our simulation, a small vortex loop in the form of an arc of a circle with ends on the wall of the channel is initially placed in the $z = 0$ plane such that its self-induced velocity is downward (i.e. in the negative $z$ direction). A maximum in the upward potential flow through the channel is assumed to occur at time $t = 0$. Each segment of the vortex loop is assumed to move with a velocity equal to the sum of the local potential flow velocity and the self-induced velocity due to the vortex itself. The self-induced velocity of each segment of the vortex loop is calculated using a local-induction approximation in which the loop is locally fit to a portion of a circle and the velocity is taken to be that of a complete vortex ring of that radius as given by Eq. (2).[29,30] The presence of the wall of the channel is taken into account only by requiring that the vortex loop meet the wall perpendicularly at its ends. The appropriateness of the local-induction approximation is discussed in Refs. 29 and 30, where it is argued that this approximation captures the essential physics in most situations. The approximation is most appropriate in situations like the present one, in which the close approach of nonadjacent segments of the vortex core does not occur and in which segments away from the ends do not come close to the wall.

The initial vortex configuration is approximated by 15 linear segments of equal length connecting 16 points, including two end points located on the wall. The vortex core is advanced in time using a 4th-order Runge-Kutta algorithm for all of the points other than the end points. At the end of each time step, the end points are simply moved to positions on the wall lying closest to the new positions of the vortex points adjacent to the end points.

Typically, extra points have to be inserted into the vortex point list as the simulation is run. This adjustment is necessary to counteract the tendency for the distance between an end point of the vortex and the point adjacent to it to become excessively large as the vortex loop expands during its evolution. The insertion is done by comparing the distance between the end point and the point adjacent to it with the distance between adjacent points near the center of the vortex loop. Whenever the former distance exceeds the latter by more than a factor of 1.5, a new point is added halfway between the end point and its neighbor. The neighboring point itself is then shifted from its current position toward the next point along the loop by 1/9$^{th}$ of the distance to that point. This simple ad-hoc point-adding algorithm is effective in keeping the vortex loops smoothly curved over their entire length for the duration of the simulation.

In this simulation, due to the presence of the length $a$ characterizing the size of the aperture, it is not possible to introduce dimensionless variables that nearly transform away the dependence of the equations of motion on $Q_0$, the amplitude of the volume rate of flow through the aperture.

However, for the case of an aperture in a nearly flat wall, i.e. for small $\eta_0$, one would expect that as the vortex travels far away from the aperture its motion would become independent of $a$. Nevertheless, this expectation does not necessarily mean that its motion will be the same as in the radial flow case. We shall see in the next section that in some situations the initially semicircular vortex loop becomes somewhat elliptical as it evolves, especially when the frequency of oscillation is close to the threshold for recapture.

### D. Results from the Oblate-Spheroidal Potential Flow Simulation

Vortex loop evolution in steady and oscillatory potential flow was studied for $\eta_0 = 0.50$ (the constricted-channel case) and $\eta_0 = 0.05$ (the case of an aperture in a nearly flat plate). The radius of the constriction was chosen to be 1.0 µm and the initial radius of the half-ring was chosen to be 0.2 µm, 1/5$^{th}$ of the radius of the constriction. The maximum volume rate of potential flow $Q_0$ was chosen to be 4.0 x 10$^{-12}$ m$^3$/s, the same value used in the radial potential flow simulation. These choices result in an initial self-induced velocity of the vortex downward that is 26% of the initial upward potential flow velocity at the mid-point of the vortex loop for $\eta_0 = 0.50$ and 30% for $\eta_0 = 0.05$. Thus the vortex is always initially swept up and out of the aperture.

For steady potential flow, our results are similar to those of Schwarz, who also considered the case $\eta_0 = 0.50$.[17-19] For oscillatory potential flow, our results are qualitatively similar to those found in the radial potential flow simulation, with some exceptions noted below. Figure 4(a) shows an oblique view of the evolution of a vortex loop for $\eta_0 = 0.50$ in steady potential flow as it crosses over the aperture and heads off toward infinity. Comparable vortex evolution in oscillatory potential flow at 400 Hz is shown in Fig. 4(b), where the total time elapsed in the simulation corresponds to one full oscillation of the flow.

Unlike vortex loops in the radial potential flow simulation, vortex loops evolving in oblate-spheroidal potential flow do not always remain exactly semicircular. Figure 5 shows one such vortex that is clearly elliptical as it leaves the vicinity of the aperture. Elliptical vorticies bend back and forth in a regular way as they propagate, exchanging their semi-major and semi-minor axes. This bending is clearly visible in the top and side views. It should be noted that the frequency of this shape-oscillation is dictated by the size and eccentricity of the vortex loop and is distinct from the oscillation frequency $f$ of the background potential flow.

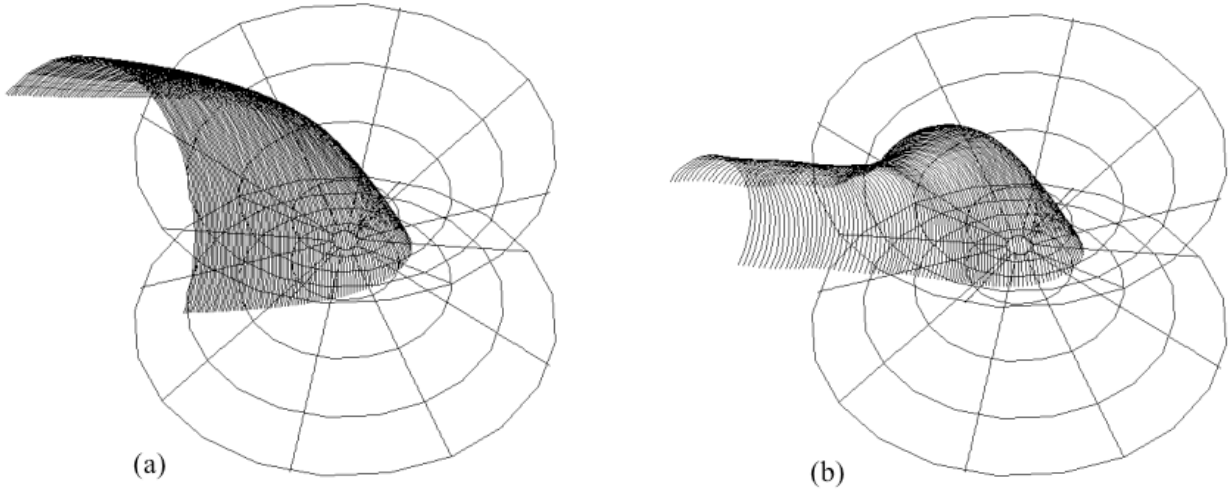

FIG. 4. Oblique views of a semicircular vortex loop evolving across an aperture in an oblate-spheroidal potential flow field with $\eta_0 = 0.50$ for steady and oscillatory potential flow, analogous to Figs. 1(b) and 1(c). (a) Vortex loop evolution in steady potential flow, showing how the vortex crosses over the aperture as it evolves. (b) Vortex evolution in oscillatory potential flow at 400 Hz, in which the vortex loop shrinks in size when the flow reverses. The total time duration shown in each plot is the same, equal to one full period of potential flow oscillation at 400 Hz.

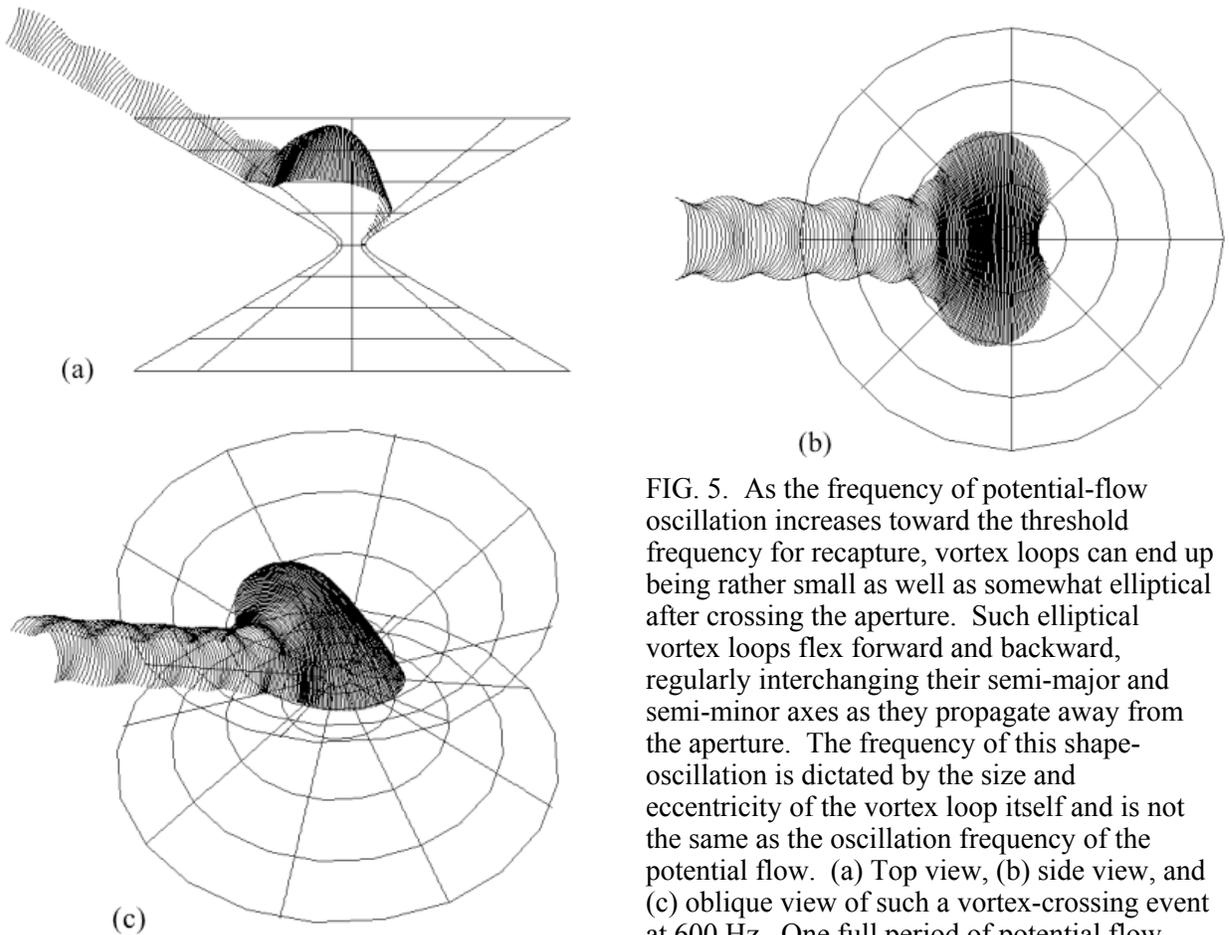

FIG. 5. As the frequency of potential-flow oscillation increases toward the threshold frequency for recapture, vortex loops can end up being rather small as well as somewhat elliptical after crossing the aperture. Such elliptical vortex loops flex forward and backward, regularly interchanging their semi-major and semi-minor axes as they propagate away from the aperture. The frequency of this shape-oscillation is dictated by the size and eccentricity of the vortex loop itself and is not the same as the oscillation frequency of the potential flow. (a) Top view, (b) side view, and (c) oblique view of such a vortex-crossing event at 600 Hz. One full period of potential flow oscillation is shown.

Figure 6 illustrates the trajectories of the peak (i.e. the mid point) of a vortex half-ring for both $\eta_0 = 0.50$ and $\eta_0 = 0.05$, in steady flow as well as for a series of frequencies of oscillatory potential flow. Arrows and zeros are shown on the traces, as was done in Fig. 2, to indicate the progress of the vortex every quarter cycle of flow oscillation. As before, the vortex eventually travels far enough away from the aperture so that the oscillation of the flow no longer affects its self-induced motion significantly. The prominent oscillations of the peak position at 600 Hz in Fig. 6(a) and at 1200 Hz in Fig. 6(b) reflect the effects of ellipticity and not of oscillations of the potential flow.

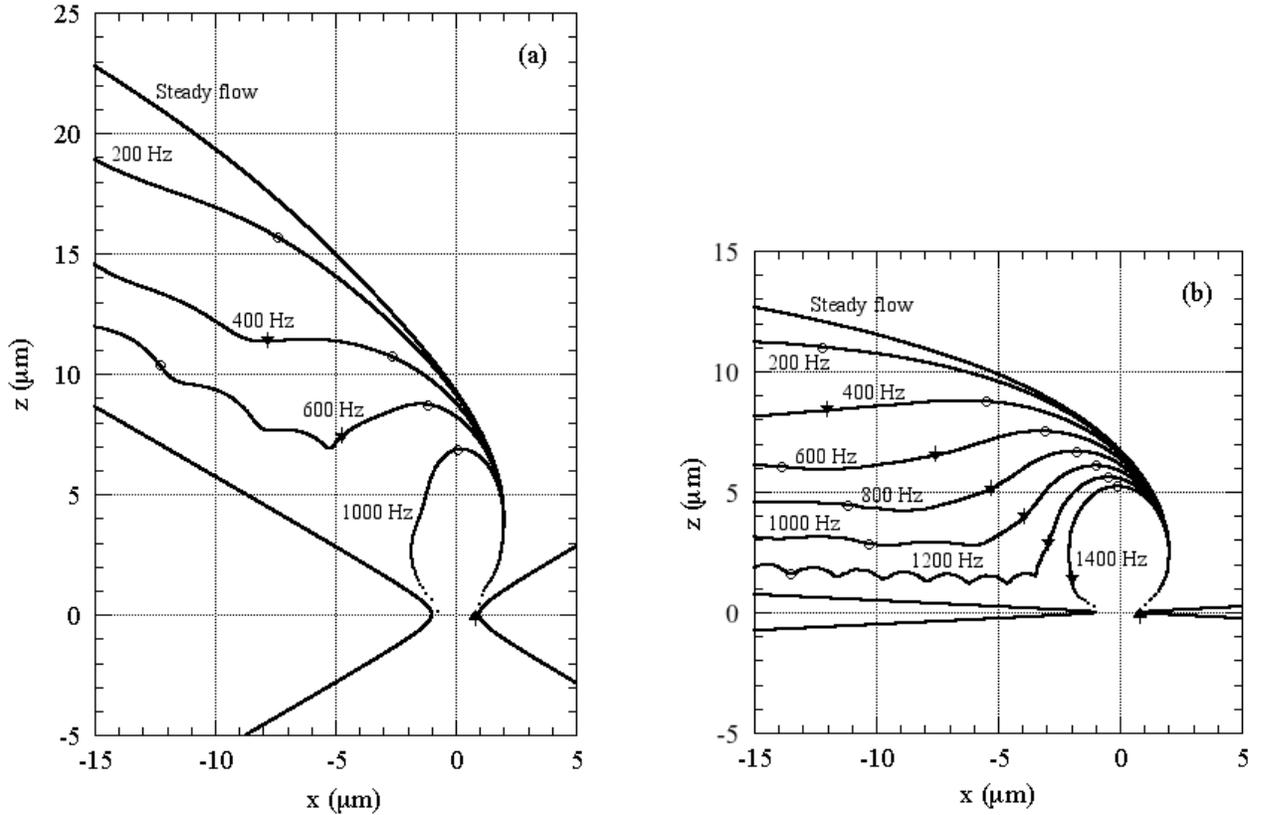

FIG. 6. Side views of the trajectories of the peak (i.e. the mid point) of a vortex half-ring evolving in oblate-spheroidal potential flow for (a) $\eta_0 = 0.50$ and (b) $\eta_0 = 0.05$. As in Fig. 2, up and down arrows and zeros on the traces indicate the progress of the vortex every quarter cycle of potential flow oscillation. The vortex trajectories for steady potential flow are also shown for comparison. The smaller the value of $\eta_0$, the more the aperture resembles a circular hole in a flat plate. Near the threshold frequency for recapture the escaping vortex loops become somewhat elliptical, resulting in "bumpy" traces on such a plot. In (b), the trace for 800 Hz has been omitted because the simulation at that frequency leads to a highly distorted vortex loop passing very close to the left-hand wall of the aperture, a situation for which the assumptions underlying our method break down.

It is difficult to calculate precisely the net amount of energy ultimately transferred from potential flow to vortex flow in the $2\pi$ phase slips in this simulation. This is because the escaping vortex loops are not simply circular arcs and also because they are bounded by a curved surface. However, it is clear from Figs. 6(a) and 6(b) that the average size of the escaping vortex goes down monotonically as the potential flow oscillation frequency goes up. Thus the amount of energy ultimately transferred from potential flow to vortex flow by such vortex-crossing events gets progressively smaller with increasing frequency. As before, there exists a threshold frequency above which vortex loops do not escape after crossing the aperture but are drawn back into the aperture when the flow reverses. This frequency lies between 700 and 1000 Hz for $\eta_0 = 0.50$ and between 1200 and 1400 Hz for $\eta_0 = 0.05$.

With this simulation it is also possible to study the ultimate fate of vortices that are drawn back into the aperture, as long as the vortex configurations remain such that nonlocal contributions and reconnections can be ignored. We have not attempted to do this systematically.

### III. CONCLUSIONS AND DISCUSSION

Both of the simulations of vortex loop evolution in oscillatory potential flow through an aperture lead to the same general conclusions. In these highly-idealized symmetrical situations, the vortex either remains nearly semicircular throughout its evolution or becomes somewhat elliptical as its motion is influenced by the oscillations of the flow. The ultimate escape size of the vortex loop, if indeed the vortex escapes at all, is smaller than the escape size in steady potential flow and decreases monotonically as the oscillation frequency of the flow increases. Thus the amount of energy dissipated from the potential flow during an individual $2\pi$ phase slip event goes down with increasing frequency. There exists a threshold frequency above which a vortex loop will not escape as it attempts to cross the aperture, but is drawn back into the aperture when the flow reverses. Such a vortex might eventually escape the vicinity of the aperture, but only after going through further evolution that we have not studied systematically.

As mentioned in the introduction, there are three experiments on which these results may have some bearing. In two experiments, in which critical velocity behavior was studied as a function of the frequency of oscillation of the flow, a transition in behavior was seen at low temperatures as the frequency was increased.[14,22-24] In the first of these experiments, studying flow through a rectangular aperture in the form of a narrow slit 5 μm by 0.3 μm in size in a 0.2-μm-thick metal foil, regular critical flow behavior was seen at frequencies of 118 Hz and below that was consistent with independent $2\pi$ phase slips, although individual slips could not be resolved.[22] However, at frequencies between 1800 and 1900 Hz and at temperatures below 1.7 K, critical velocity behavior was seen that involved large, irregular energy-loss events, each involving many times the energy loss of an individual 2π phase slip.

In the second experiment, a similar phenomenon was observed and studied in more detail, using a 2-μm by 2-μm square aperture in a 0.1-μm-thick metal foil.[14,23,24] In this experiment, individual 2π phase slips were resolved and observed to dominate the critical behavior at frequencies of 80 Hz and below. As the frequency was raised, large, irregular energy-loss events were observed to occur above 500 Hz at temperatures below 1.1 K. Each of these events was spread over tens of potential-flow reversals, rather than occurring within a single half-cycle of oscillation, and dissipated on the order of hundreds of times the energy that would have been transferred to a single vortex loop crossing the aperture at the maximum rate of the background potential flow.

Our simulations, whose parameters were chosen to be representative of this second experiment, suggest that these events cannot simply be explained as being due to a large number of individual, independent vortex-loop-crossing events. At the frequencies involved, single vortex-loop events in the simulations become increasingly ineffective at dissipating energy from potential flow, and many loops would have had to be present at the same time. On the contrary, some new dissipation mechanism seems to be involved. The approximate agreement between the observed onset frequency and the threshold frequencies in the simulations suggests that the new mechanism may be triggered by the drawback of vortices into the aperture.

It has been suggested that under certain conditions, the critical velocity for dissipation in steady potential flow through apertures or narrow channels may be due to the establishment of a "vortex mill", a process in which dissipation results from the continuing production of vortex line without the need for nucleation to occur repeatedly.[31,32] Such a process has been invoked to explain the relatively low and temperature-independent critical velocities that are often seen in long channels ~10 μm and larger in diameter. An example of one such vortex mill has been simulated by Schwarz for steady oblate-spheroidal potential flow through a constriction.[17-19] In his simulation, one end of a vortex filament ending at the wall in the aperture region travels around the aperture repeatedly, while vortex reconnections produce a succession of closed-loop vortices that carry energy away from the potential flow through aperture. However, we do not have a clear picture of how such a process might be initiated in oscillatory potential flow by the drawback of vorticity into the aperture. Nor is it clear whether such a mill is capable of continuing to dissipate energy throughout a series of flow reversals, as is observed experimentally.

The third experiment involved a study of individual $2\pi$ phase slips at the critical velocity of superfluid flow through a 2-μm by 2-μm square aperture in a 0.1-μm-thick metal foil very similar to the aperture used in the second experiment above, although the critical velocity in this experiment was found to be much larger than in the second experiment.[13] Within considerable experimental uncertainty, the energy loss per $2\pi$ phase slip relative to the loss expected for steady flow was observed to have a temperature dependence consistent with the effects of oscillatory flow predicted by the simple radial-flow model of Sections II A and II B. These measurements were carried out at frequencies from 19 Hz down to 10 Hz, with critical values of $Q_0$ of $40 \times 10^{-12}$ m$^3$/s and below. The fact that the effects of the variation of the flow with time were expected to be visible at frequencies much lower than those of Fig. 2 is consistent with the larger values of $Q_0$ here than for Fig. 2 and the scaling arguments of Section II B.

In closing a discussion of experiments, it should be mentioned that some time ago, Hess and coworkers observed an influence of a bias current on the critical velocity behavior of an aperture that was attributed to the drawback of vorticity.[33]

It would be of considerable interest to extend simulations of the type discussed here to less symmetrical initial configurations and to pursue the fate of vortex loops drawn back into the aperture. This would almost-certainly require going beyond the local-induction approximation, allowing for vortex reconnections, and treating the effects of boundaries more accurately.

## ACKNOWLEDGEMENTS

We would like to acknowledge the major contribution of C. L. Zimmermann, who wrote the original code for the oblate-spheroidal potential flow simulation. This work was supported by the National Science Foundation under grants DMR 90-02890, 94-03522, and 96-31703, and by Luther College and the University of Minnesota – Morris, where many of the simulations by the first two authors were run. The University of Minnesota's UROP program is also gratefully acknowledged.